\documentclass[12pt,preprint]{aastex}

\def\lessim{\mathrel{\hbox{\rlap{\hbox{\lower4pt\hbox{$\sim$}}}\hbox{$<$}}}}
\def\grtsim{\mathrel{\hbox{\rlap{\hbox{\lower4pt\hbox{$\sim$}}}\hbox{$>$}}}}

\shorttitle{LMC Novae}
\shortauthors{Shafter}

\begin{document}


\title{Photometric and Spectroscopic Properties of Novae in the Large Magellanic Cloud}


\author{A. W. Shafter}
\affil{Department of Astronomy, San Diego State University, San Diego, CA 92182, USA}



\begin{abstract}
The photometric and spectroscopic properties of the 43 known LMC nova
candidates are summarized and reviewed. Of these,
photometric data sufficient to establish decline rates are available
for 29 novae, while spectroscopic data sufficient to establish the
spectroscopic classes are available for 18 systems. Half of the
18 novae belong to the Fe~II class, with the remaining nine
belonging to either the He/N or the Fe~IIb classes.
As seen in previous nova studies of M31 and M33, the He/N and Fe~IIb
novae have on average faster photometric developments than do their
Fe~II counterparts. Overall, the available photometry confirms
earlier studies, and shows conclusively that LMC novae have faster
rates of decline than do novae in the Galaxy and M31. It appears that
the increased fraction of faster, He/N and Fe~IIb novae observed in
the LMC compared with M31 is almost certainly the result of differences
in the underlying stellar population between the two galaxies. We propose
that the younger population seen in the LMC compared with M31's bulge
(where most of the novae are found)
produces progenitor binaries with higher average white dwarf masses.
The higher mean white dwarf mass not only produces a larger fraction of fast,
He/N novae compared with M31, but also results in a relatively large
recurrent nova population.

\end{abstract}




\section{Introduction}

Over the past two decades it has become widely accepted that the observed
properties of novae -- the peak luminosity, the rate of decline, and
the character of the post-eruption spectrum are strongly affected
by the underlying stellar population
\citep[e.g., see][and references therein]{sha08}.
In studies of Galactic novae, \citet{due90} proposed the idea of two distinct
populations of novae: a relatively young population of rapidly-fading novae
found in the solar neighborhood, the so-called ``disk" novae, and
a population of slower developing ``bulge novae", which are concentrated
towards the Galactic center. At about the same time,
\citet{wil92} argued that the spectra of Galactic novae
could be divided into one of two principal spectroscopic classes.
Novae that display prominent Fe~II emission (the ``Fe~II" novae)
are characterized by P Cygni line profiles, a slow photometric
evolution, lower expansion velocities, and lower levels of ionization,
compared to novae with strong lines of He and N (the ``He/N" novae).
The spectroscopic types are believed to be related to fundamental properties
of the progenitor binary such as the white dwarf mass, and are possibly
dependent on the underlying stellar population \citep[e.g.,][]{del98}.

In order to explore the effect of stellar population on nova properties,
\citet{sha11a, sha11b, sha12} studied the spectroscopic properties of novae
in the Local Group galaxies M31 and M33. In M31 (and the Galaxy), Fe~II novae
make up $\sim$80\% of novae that have been observed spectroscopically,
while in M33, preliminary evidence suggests that Fe~II novae make up
a significantly smaller fraction of novae.
Among the 91 M31 novae with known spectroscopic class studied by \citet{sha11b},
no clear dependence of a nova's spectroscopic type
on spatial position (and presumably stellar population) in M31 was found.
This result may be misleading, however, since the high inclination of
M31's disk to our line of sight makes
it difficult to assign a given nova an unambiguous spatial position
within the galaxy,
particularly near the apparent center of M31 where the foreground
disk is superimposed on the galactic bulge. Despite this problem,
the available photometric data
did suggest that rapidly declining novae have a 
slightly more extended spatial distribution, as expected for a
disk population.
Given the difficulties in isolating stellar populations within
a given galaxy, it would appear that a more promising approach
to exploring differences in nova populations
would be to compare nova properties
between galaxies with differing Hubble types where population synthesis
models have predicted they should vary \citep[e.g.,][]{yun97}.

In this paper, we review all available photometric and spectroscopic
observations of novae in the Large Magellanic Cloud (LMC) up to
the end of 2012. We then
compare their properties with those of novae in M31, M33 and the Galaxy
with the goal of furthering our understanding
of how nova properties vary with stellar population.

\section{Spectroscopic Classification of Novae}

Nova outbursts are the result of a thermonuclear runaway (TNR)
in the degenerate surface layers of an accreting
white dwarf in the progenitor binary. The TNR
ejects some or all of the accreted material, and in some cases may
dredge up material from the white dwarf itself.
Spectra of novae shortly after eruption (days to weeks)
are dominated by prominent Balmer emission lines.
In addition, the immediate post-eruption
spectra often display prominent emission lines
of either Fe~II or He and N in various stages
of ionization \citep{wil92}. The former group, which represents
$\sim$80\% of Galactic and M31 novae \citep{sha07,sha11b},
shows spectra dominated by low excitation,
relatively narrow lines (FWHM H$\alpha$ $\lessim$ 2000~km~s$^{-1}$)
that are flanked by P Cygni absorption components
characteristic of an expanding, optically thick gas.
The remaining $\sim20$\% of novae, the ``He/N" systems, display
higher excitation emission lines that are generally
broader (FWHM of H$\alpha$ $\grtsim$ 2500~km/s) that are characterized
by more rectangular, castellated or flat-topped profiles.
A small fraction of novae, which appear to have
characteristics of both classes, are referred to as either
``hybrid" or broad-lined Fe~II (Fe~IIb) novae. These systems appear
similar to Fe~II novae shortly after eruption, but their
emission lines are broader. Later, they may
evolve to display a classical He/N spectrum, but
have not been seen to evolve into typical narrow-lined Fe~II novae.
Empirically, spectral classifications made within the first few weeks after
eruption seem to be robust; and, with the exception of
the hybrid novae, do not appear to be sensitive to the precise
phase in the evolution of the outburst. That said, additional
time-resolved spectroscopic
observations of novae throughout their outburst development will be required
before the stability of the spectral classification can be fully assessed.

The properties of the progenitor binary that determine the nova's
spectroscopic class are not completely understood.
While it seems clear that the He/N spectrum
must be produced in a relatively low-mass ejecta consisting of an
optically-thin shell of gas that is
ejected at high velocity in the eruption \citep{wil92,wil12}, the origin of
the Fe~II emission spectrum is more uncertain. The conventional explanation
has been that Fe~II spectra are produced in the optically-thick wind that
results from the expulsion of massive ejecta from the white dwarf's surface
\citep{wil92}.
More recently, however, \citet{wil12}
has suggested that the Fe~II emission spectrum
is produced predominately in gas stripped from the secondary star during
the eruption.
In this picture, nova binaries with large mass ratios, $q~(=~M_{sec}/M_{wd})$,
where the secondary star subtends a relatively large angle as seen
from the white dwarf,
would be expected to be more likely to produce Fe~II spectra.

Regardless of whether the emission originates in the wind or in material
stripped from the secondary star, Fe~II novae are likely to
harbor low-mass white dwarfs.
Models show that the accreted mass required
to trigger a TNR (the ignition mass)
is primarily a function of the white dwarf mass and temperature, with the
latter being strongly influenced by the rate of accretion onto the
white dwarf's surface. Thus, novae harboring low-mass
white dwarfs accreting slowly will have the largest accreted
masses and the longest recurrence times between eruptions.
Assuming the ejected mass is proportional to the accreted mass,
such systems will be more likely to produce the large ejected masses
required to produce an optically-thick wind. In addition,
the high mass ratios favored in systems with low mass white dwarfs
will act to increase the interaction of this wind
with the secondary star, further
facilitating the production of an Fe~II spectrum.
In He/N novae, on the other hand,
it is thought that a relatively small amount of gas is ejected quickly
from a relatively massive white dwarf, with
little contribution from a wind. In this case, the spectrum will
be dominated by emission from a high-velocity shell ionized by
the hot white dwarf resulting in a nova
of the He/N spectroscopic class. Thus, it appears that a study
of the spectroscopic classes of novae can shed light on the
white dwarf mass distribution in a nova population.

\section{Novae in the LMC}

A total of 43 suspected nova eruptions have been observed in the LMC
since the first nova was discovered in 1926 \citep{luy26}\footnote
{\tt See http://www.mpe.mpg.de/\~{}m31novae/opt/lmc/index.php for
a compilation of positions, discovery magnitudes and dates.}.
Five of these recorded outbursts are uncertain
(LMCN 1952, 1966, 1996, 1998 and 1999), and another
three (LMCN 1990-02a, 2004-10a, and 2009-02a)
almost certainly represent the second outburst of a previously
known system, and are therefore recurrent novae (RNe).
Thus, the likely number of independent nova systems
identified in the LMC as of the end of 2012 is 35.
Given the challenges associated with obtaining spectra of transient
sources like novae, it is perhaps surprising that spectroscopic
observations sufficient to determine the spectroscopic class
is available for more than half of these novae.
Specifically, an analysis of published spectroscopic data
has allowed us to assign tentative spectroscopic classes:
Fe~II, Fe~IIb (hybrid), or He/N, for a total of 22 LMC novae.
Of these 22, 17 spectroscopic classes are reasonably secure.
A summary of the properties of all known LMC nova candidates, including
their spectroscopic classes where known, is given in Table~\ref{novatable},
with the spatial distribution of the novae shown in Figure~1.
Below we summarize the spectroscopic
properties of the 20 LMC novae for which
spectroscopic classifications are possible. When available,
we have also included estimates of the magnitude reached at peak brightness
and the rate of decline, $\nu$. Often the rate of decline is parameterized
in terms of the number of days a nova takes
to fade either two or three magnitudes from maximum
light, $t_2$ or $t_3$.

\noindent
{\bf $\bullet$ LMCN 1970-03a:}
LMCN 1970-03a was discovered by \citet{mag70} on an objective
prism plate taken on 1970 March 8.2 UT.
\citet{mac70} describes that objective prism spectrum as follows:
``H is very bright, flat-topped, and broad and
faint, broad Fe~II emission at $\lambda$4924 and $\lambda$5018
is present."
Although,
line widths are not given, based on
MacConnell's description,
it appears that LMCN 1970-03a is most likely
a broad-lined Fe~II nova. The nova was discovered at $V\sim12$,
unfortunately, with no light curve information available.

\noindent
{\bf $\bullet$ LMCN 1970-11a:}
LMCN 1970-11a was discovered in decline, with maximum light not covered
by available observations. Based on extrapolation of available data,
\citet{cap90} estimate that LMCN 1970-11a reached $V\sim10.5-11.0$.
Available observations showed that the nova faded quite rapidly,
with \citet{gra71} estimating a decline rate of 0.25 mag per day.
Spectra obtained by \citet{hav72} starting 9 days after the
estimated date of maximum light on 1970 October 30
shows the nova to be a member of the Fe~II
spectroscopic class. They deduce a shell velocity of $-1560$~km~s$^{-1}$,
and estimate that the nova may have reached $V=10.8$ at maximum light.

\noindent
{\bf $\bullet$ LMCN 1977-03a:}
LMCN 1977-03a was discovered by \citet{gra77} on 1977 March 12.
According to \citet{can77} the nova reached $V = 10.7$ and suffered
a reddening of $E(B-V) = 0.11$. If we adopt
a ratio of total-to-selective extinction of $R=3.2$, and a distance
modulus, $\mu_o = 18.50$ \citep{fre01}, we find an absolute magnitude at maximum
light, $M_V = -8.2$. A subsequent analysis of the nova by
\citet{can81} found that $t_2(V) = 11$~d and $t_3(V) = 21$~d, respectively.
These authors also reported spectroscopic observations near the time of
maximum light that are consistent
with the identification of LMCN 1977-03a as an Fe~II system.

\noindent
{\bf $\bullet$ LMCN 1978-03a:}
LMCN 1978-03a was discovered by \citet{gra78a} on 1978 March 29 UT
at $V = 12$, and later confirmed by \citet{gra78b}. An objective prism
spectrum obtained at the time of discovery
was described by \citet{gra79}, and shows the Balmer series in emission.
An inspection of the published spectrogram
suggests that the FWZI of H$\beta$ is $\sim2400$ km~s$^{-1}$. Additionally,
there is a hint that N~III $\lambda4640$ may be present. A pre-maximum
objective prism spectrum was also obtained
by \cite{gra79} on 1978 March 17 UT as part of
routine monitoring of the LMC. The spectrum showed narrow Balmer absorption,
and may indicate that the true maximum of the nova occurred much earlier
that the reported discovery on March 29. An extrapolation of the
light curve from March 29 suggests that the nova could have reached
$V\sim9.5-10$ at maximum light and faded with a rate of
$\sim$0.5 mag per day \citep{gra78b}. We estimate that the spectral
classification occurred $\sim$10 days after maximum light.

\noindent
{\bf $\bullet$ LMCN 1981-09a:}
LMCN 1981-09a was discovered by J. Maza on 1981 September 30.371~UT
at $m_{pg}\sim12$,
and later confirmed by P. Jekabsons on October 6.85~UT.
A day later, on October 7.64~UT, A. A. Page found that the nova
had brightened to $m_{pg}=11.8$.
An objective prism spectrum obtained by
H. Duerbeck near the time of maximum light
showed very strong and broad H$\alpha$ emission
(FWHM = 3800~km~s$^{-1}$), along with Fe~II and N~II emission.
These data suggest that LMCN 1981-09a is a Fe~IIb or hybrid nova.
A summary of all early observations can be found in \citet{maz81}.
Unfortunately, little light curve information exists, and
a reliable estimate of the fade rate is not available.

\noindent
{\bf $\bullet$ LMCN 1988-03a:}
LMCN 1988-03a was discovered by G. Garradd on 1988 March 21.484~UT
at $V\simeq11.4$ \citep{gar88}.
A series
of spectroscopic observations made in the first month following
discovery revealed relatively
narrow Balmer and Fe~II emission features \citep{sch98,dre90}, indicating
that the nova is a member of the Fe~II class. An
analysis of the light curve by \citet{hea04} indicated that the nova
reached $V=11.2\pm0.3$ and declined moderately rapidly,
with $t_2(V)=22.5\pm4$.

\noindent
{\bf $\bullet$ LMC 1988-10a:}
The second nova observed in 1988 was also discovered by G. Garradd
on 1988 October 12.48~UT at $V=11.3$ \citep{mcn88},
before brightening to $V=10.3$
on October 13.75~UT \citep{sek89}. Extensive spectroscopic
observations by \cite{sek89} revealed prominent He, N, and Balmer
emission lines (FWZI $\sim6000$~km~s$^{-1}$) characteristic of the He/N novae.
An analysis of the light curve, also by \citet{sek89}, established
that the nova faded relatively rapidly with $t_2(V)$ and $t_3(V)$ values
of 5 and 10 days, respectively.

\noindent
{\bf $\bullet$ LMCN 1990-01a:}
LMCN 1990-01a was discovered on 1990 January 16.47 UT
at magnitude 11.5 \citep{mcn90}.
Spectra obtained by \citet{dop90} between one and two weeks post-discovery
reveal broad (FWHM $\sim5600$~km~s$^{-1}$),
flat-topped Balmer and He~I emission early on followed by
increasing He~II and [Ne~III] lines by January 30. These data establish
LMCN 1990-01a as an ONe nova, and a member of the He/N spectroscopic class.
An analysis of the light curve by \citet{lil05a} suggests that the nova reached
$V=9.7$ at maximum light (although maximum was not observed directly) and
that the nova faded rapidly at a rate of 0.59 mag per day.
With
an estimated reddening, $E(B-V) = 0.22\pm0.07$ \citep{van99}, the nova
may have reached $M_V=-9.6$.

\noindent
{\bf $\bullet$ LMCN 1990-02a:}
LMCN 1990-02a was discovered at the position of LMCN 1968-12a
by Liller on 1990 February 14.1~UT at $V=11.2$,
making it the first recurrent nova to be recognized in the LMC
\citep{sek90a,sho91,wil90}. Spectroscopic observations by \citet{sek90b}
and \citet{sho91} obtained approximately a week post discovery
established that the nova was a member of the He/N
spectroscopic class. The nova was observed to fade rapidly, dropping
and estimate 4.5 magnitudes in the week since discovery. \citet{lil05a}
estimate that the nova likely reached $V=10.2$ and faded at a rate of $\sim$0.59
mag per day.

\noindent
{\bf $\bullet$ LMCN 1991-04a:}
LMCN 1991-04a was discovered on the rise to maximum \citep{lil91}.
The nova reached $V\simeq9.0$ on 1991 April 24~UT,
making it the brightest nova ever observed in the LMC \citep{sho91,del91a}.
After reaching maximum, the nova was observed to fade relatively rapidly
with $t_2(V)$ and $t_3(V)$ of 6 and 8 days, respectively \citep{sho91}.
A spectrum obtained more than 2 weeks after maximum light
was that of an Fe~II nova, but with relatively broad
Balmer emission lines (FWHM H$\alpha$ = 2500~km~s$^{-1}$), along
with a prominent CIII/NIII blend at $\lambda$ 464.5 nm \citep{del91b}.
Thus, it appears
that LMCN 1991-04a may be a member of the Fe~IIb spectroscopic class.

\noindent
{\bf $\bullet$ LMCN 1992-11a:}
LMCN 1992-11a was discovered by \citet{lil92} at $R=10.7$
on 1992 November 11.21~UT.
Subsequent photometry showed that the nova reached $V=10.2\pm0.2$ and faded
relatively quickly with $t_2(V)=6.9\pm1.1$~d and $t_3(V)=13.7\pm1.6$~d
\citep{hea04}.
Spectroscopic observations by \citet{del92} and \citet{due92}
near maximum light clearly establish
that the nova is a member of the Fe~II spectroscopic class.

\noindent
{\bf $\bullet$ LMCN 1995-02a:}
LMCN 1995-02a was discovered at $m=10.7$ by \citet{lil95}
on 1995 March 2.11~UT.
An analysis of the light curve by \citet{hea04} revealed that the nova
reached $V=10.35\pm0.5$ and subsequently faded with characteristic
times of $t_2(V)=11\pm3$~d and $t_3(V)=19.6\pm3.2$. Limited spectroscopic
observations by \citet{del95b}
obtained at or near maximum light suggests that the nova
was likely a member of the Fe~II class.

\noindent
{\bf $\bullet$ LMCN 2000-07a:}
LMCN 2000-07a was discovered by \citet{lil00} on 2000 July 13.4~UT.
Although maximum light was not covered,
\cite{hea04} estimate that the nova reached $V=10.7\pm0.5$ and that it
subsequently faded with $t_2(V)$ and $t_3(V)$ times of 8 and 20 days,
respectively. Photometry by \citet{gre03} resulted in similar values for
$t_2$ and $t_3$.
Spectroscopy obtained approximately 3 days after discovery
by \citet{due00} established that the nova
was a member of the Fe~II spectroscopic class.

\noindent
{\bf $\bullet$ LMCN 2002-02a:}
LMCN 2002-02a was yet another nova discovered by \citet{lil02}. According
to his discovery images, the nova reached $V = 10.5$ on 2002 March 3.066~UT.
\citet{lil05a} estimate that the nova reached $V=10.1$ at maximum light.
Subsequent photometric and spectroscopic follow-up observations by
\cite{mas05} obtained approximately a week post discovery
established that LMCN 2002-02a was a member of the
Fe~II spectroscopic class, and that it faded with a $t_2=12$~d.
\citet{sub02} give $t_3=23$~d, suggesting a slightly slower rate of decline,
possibly based on the \citet{lil05a} estimate of $\nu=0.1$~d$^{-1}$.

\noindent
{\bf $\bullet$ LMCN 2003-06a:}
LMCN 2003-06a was discovered by \citet{lil03} at $V=11.0$ on 2003
June 19~UT. Little is
known about the spectroscopic properties of this nova, but H. Bond
reported that the emission lines were very broad \citep{lil03}. An
analysis of a blue spectrum ($\lambda$3900--4500 \AA) obtained 11
days post discovery
kindly provided by H. Bond (2012, private
communication) shows that FWHM (H$\gamma$) = 3600~km~s$^{-1}$. The nova
appears to be either an Fe~IIb, or more likely an He/N system.
\citet{lil05a} estimate a decline rate of 0.25 mag per day.

\noindent
{\bf $\bullet$ LMCN 2004-10a:}
LMCN 2004-10a was discovered by \citet{lil04} on 2004 October 20.193~UT near the
position of LMCN 1937-11a (YY Dor). Spectroscopic observations by
\citet{bon04} and by \citet{mas04} during the first week post discovery
clearly establish the nova as belonging to the He/N class,
consistent with its identification as a recurrent nova.
Limited photometric observations suggest that the nova reached
$V\sim10.9$, and faded at a rate of $\sim$0.17 mag per day \citep{lil05a}.

\noindent
{\bf $\bullet$ LMCN 2005-09a:}
LMCN 2005-09a was unusual in that it was identified on 2006 July 18~UT
via its X-ray emission approximately 10 months
after eruption \citep{rea09}. Analysis
of archival photometry from the All Sky Automated Survey \citep{pom02}
revealed that the nova erupted sometime between 2005 September 18 and 30,
when it was observed at $V\sim12$. The very limited photometric
data suggested that the nova faded relatively rapidly, with $t_2(V)\sim8$~d.
The only spectroscopic data were obtained by \citet{rea09} more than
a year after eruption. Although a reliable
spectroscopic classification is not
possible this long after eruption, \citet{rea09} have speculated that
the nova might have been an Fe~II system.

\noindent
{\bf $\bullet$ LMCN 2005-11a:}
LMCN 2005-11a was discovered by \citet{lil05b} on 2005 November 22.065~UT
at $m\sim12.8$. An analysis of the light curve by \citet{lil07}
shows that the nova likely reached $V=11.5$ and then
faded very slowly with $t_2(V)=63$~d and $t_3(V)=94$~d.
Spectroscopic observations by \citet{wal05, wal12}
starting 2 days post discovery show the object to be
a member of the Fe~II spectroscopic class.

\noindent
{\bf $\bullet$ LMCN 2009-02a:}
LMCN 2009-02a was discovered by \citet{lil09a}
on 2009 February 5.067~UT at $m\sim10.6$. The eruption likely represents a
recurrence of the poorly observed nova, LMCN 1971-08a
(Bode et al. 2013, in preparation). Spectroscopic
observations by \citet{ori09} are consistent with the classification of the
nova as a member of the He/N class. Limited photometric observations suggest
that the nova faded rapidly. Based on the $V$-band photometry of
\citet{wal12}, we estimate $t_3\sim8$~d and $\nu\sim0.37$~d$^{-1}$.

\noindent
{\bf $\bullet$ LMCN 2009-05a:}
LMCN 2009-05a was discovered at $m\sim12.1$
on 2009 May 4.994 UT \citet{lil09b}. Photometric and spectroscopic
observations by \citet{wal12} starting 3 days post discovery
show that the object was
a slowly fading Fe~II nova. Based on the \citet{wal12} $V$-band
photometry, we estimate $t_3\sim80$~d and $\nu\sim0.037$~d$^{-1}$.

\noindent
{\bf $\bullet$ LMCN 2012-03a:}
LMCN 2012-03a was discovered by J. Seach on 2012 March 26.397 UT
at $m=10.7$ \citep{lil12}. The nova was observed
to fade very rapidly after discovery, with \citet{wal12}
estimating $t_2\sim1.1$~d and $t_3\sim2.1$~d. However, the
available observations do not provide tight constraints on the time of
maximum light (and thus maximum brightness), so these estimates are uncertain.
Spectroscopic observations
on March 27.0~UT shortly after discovery reveled the object to be
a member of the He/N spectroscopic class \citep{pri12}.

\noindent
{\bf $\bullet$ LMCN 2012-10a:}
LMCN 2012-10a was discovered on 2012 November 04.358 UT by \citet{wyr12}
as part of the OGLE-IV search for transients in the LMC. An examination
earlier data shows that the nova reached maximum light between
2012 October 22.374 and 25.310 UT. The October 25.310 UT image showed
the nova saturated, with an $I$-band magnitude between 11 and 12.
\citet{wyr12} estimate $t_2(I)$ and $t_3(I)$ times of 10 and 15 days,
respectively.
Subsequent spectroscopy by \citet{wal12} starting $\sim$1 week post maximum
reveals the nova to be a likely member of the He/N spectroscopic
class.

\subsection{Global Photometric Properties}

It has long been recognized that a nova's luminosity
at maximum light and its rate of decline are
correlated. The resulting Maximum Magnitude versus Rate of Decline
(MMRD) relation was first applied to Galactic novae
by \citet{mcl45}. Since that time, there have been numerous
characterizations the MMRD relation for Galactic
\citep[e.g.,][]{coh85, dow00} and extragalactic nova populations
\citep[e.g.,][]{cap90, del95a, dar06}. A common finding of all of
these studies is that there appears to be considerable intrinsic scatter in the
MMRD relation that cannot be explained solely by observational uncertainties.
A recent study by \citet{kas11} has revealed a number of apparently
faint, yet relatively rapidly fading novae in M31, which has caused
the authors to question whether an MMRD relation is justified at all.
In order to re-examine the MMRD relation for the LMC, we have
reviewed available photometric observations and
produced estimates of maximum magnitudes
and rates of decline for the 29 nova outbursts summarized in
Table~\ref{lcparam}. Values of the absolute magnitude at maximum
have been estimated by adopting a distance, $\mu_{LMC}=18.50$ \citep{fre01},
and, when unknown,
assuming a reddening $E(B-V)=0.12$ \citep{ima07}. When not reported directly,
decline rates, $\nu$, have been estimated from
$\nu=2/t_2$ or $\nu=3/t_3$. If both $t_2$ and $t_3$ are available, an
average of the resulting decline rates is adopted.

In Figure~\ref{fig2}
we have plotted the MMRD relation for LMC novae based on our estimate of the
absolute $V$ magnitudes at maximum light and the measured rates of decline
(parameterized as log~$100\nu$). For most novae,
peak brightness was not well covered by observations, making both
the magnitude at maximum light and the fade rate uncertain.
Given these unknown errors, we have
not attempted to apply
any corrections to convert observations in the photographic or unfiltered
``white light" bandpasses to the visual band, nor have we
included formal error bars when plotting the points.
Despite the considerable observational uncertainties,
a general MMRD relation is apparent,
with the more luminous novae appearing to generally fade the fastest.
A linear least-squares fit to the data (excluding the three lower limits)
gives the following relation:

\begin{equation}
M_V(\mathrm{max}) = - (1.52\pm0.24)~\rm{log}~100\nu - 6.27\pm0.32.
\end{equation}

\noindent
For comparison, we have also plotted the corresponding relation for M31,

\begin{equation}
M_V(\mathrm{max}) = - (1.70\pm0.08)~\rm{log}~100\nu - 5.87\pm0.10,
\end{equation}

\noindent
found by \cite{sha11b}, which, in light of the considerable uncertainties,
appears consistent with our LMC result.

Although there is some overlap in fade rates of He/N and Fe~II
novae, as found in M31 \citep{sha11b} and M33 \citep{sha12},
it is clear that the He/N novae are generally ``faster" than
their Fe~II counterparts. Of the 9 novae with $t_3$ times less
than or equal to 10~days (log~$100\nu \grtsim 1.5$),
none are confirmed Fe~II novae;
five are He/N or suspected He/N, one is a possible Fe~II nova,
and three are of unknown spectroscopic type. Conversely,
the five slowest novae with known spectroscopic type are all
Fe~II systems. It should be noted here that Fe~II systems
are not confined to the slowest or least luminous novae.
Indeed, there appears to be a rare class of very luminous, and slowly
rising Fe~II novae, one of which is LMCN 1991-04a, which we have argued is
likely a member of the Fe~IIb class. Two other notable examples of
highly luminous Fe~II novae with somewhat narrower emission lines are
M31N 2007-11d \citep{sha09} and SN~2010U \citep{cze13}.

It has been argued for some time that the novae in the LMC
generally faded from maximum light more quickly than novae in the
Galaxy or in M31~\citep[e.g.,][]{del93}.
In Figure~\ref{fig3}, we compare the fade rates for our complete
sample of LMC novae with those available for the Galaxy
and M31 from Tables~5 and~6 of \citet{dow00} and
\citet{sha11b}, respectively.
In agreement with earlier results,
it is clear from Figure~\ref{fig3} that the LMC novae are considerably
``faster" than their Galactic and M31 counterparts.

\subsection{Global Spectroscopic Properties}

Of the 18 novae listed in Table~\ref{novatable} having sufficient
data for a firm spectroscopic class to be assigned, half are members
of the Fe~II spectroscopic class, with the other half being members
of the He/N or Fe~IIb classes\footnote{Seven are He/N class, with just
one Fe~IIb (LMCN 1981-09a). LMCN 2003-06a is likely an He/N system,
but the limited spectral coverage does not allow an Fe~IIb
classification to be ruled out.}. For an additional three
novae sufficient data exist to assign a tentative spectroscopic class,
with LMCN 1978-03a a likely Fe~II system, and 1970-03a and
1991-04a likely being broad-lined Fe~II (Fe~IIb) novae.
LMCN 2005-09a has been tentatively classed as an Fe~II system
by \citet{rea09} based on a single spectrum taken
more than a year post eruption.
In reality, the spectroscopic class of this nova is unknown.
The fraction of Fe~II novae in the LMC
($\sim50\%$), like that found in M33 \citep{sha12},
is significantly lower than the fraction
($\sim 80\%$) seen in M31 and the Galaxy
\citep{sha11b, sha07}.

To further explore the spectroscopic properties of the LMC novae,
in Table~\ref{balmerline}
we have collected all published measurements of the H$\alpha$
emission-line widths (which reflect the expansion velocities of the
ejected gas where the lines are formed).
As is now well established from prior spectroscopic surveys of novae,
the emission-line widths are strongly correlated with spectroscopic class.
As in previous surveys of novae in
M31 and M33 \citep{sha11b,sha12},
the novae belonging to the He/N class are characterized by H$\alpha$
FWHM~$>2500$~km~s$^{-1}$, while the Fe~II systems have
FWHM values typically less than 2000~km~s$^{-1}$.
It is thought that the Balmer emission lines associated with the He/N novae,
with their broad, rectangular, and flat-topped profiles,
are formed in discrete, optically-thin shells that are
ejected at relatively high velocity
from near the white dwarf's surface at the onset of eruption.

Not surprisingly, the
emission-line widths (expansion velocities) are also clearly
correlated with the rate at which the nova fades in brightness.
The low mass ejecta associated with rapidly expanding shells
are expected to become optically thin more quickly than in more massive
ejecta, resulting is faster rates of decline.
For the LMC novae, Figure~\ref{fig4} shows the relationship
between the fade rate and the FWHM of H$\alpha$.
The dashed line shows the best-fit linear relation given by
\begin{equation}
\mathrm{log}~100\nu~=~(1.01\pm0.25)~\mathrm{log~FWHM_{H\alpha}}~-~2.04\pm0.85,
\end{equation}
\noindent
while the dotted line shows a corresponding relation given by
\begin{equation}
\mathrm{log}~100\nu~=~(1.05\pm0.34)~\mathrm{log~FWHM_{H\alpha}}~-~2.50\pm1.08,
\end{equation}
\noindent
for a sample of M31 novae from \citet{sha11b}.
In addition to highlighting the
sharp distinction in line widths between the Fe~II and the He/N
novae, the figure clearly shows that novae with broad emission-line widths
fade the fastest as expected. The He/N, which all
have a FWHM$_{\mathrm{H}\alpha}>2500$~km~s$^{-1}$, have a mean fade rate
$<$$\nu(\mathrm{He/N)}$$>$~=~($0.54\pm0.45$)~d$^{-1}$,
while the Fe~II novae have a mean
fade rate $<$$\nu(\mathrm{FeII})$$>$~=~($0.15\pm0.09$)~d$^{-1}$.
It is unclear whether the slight offset between the LMC and M31
relations given by Equations~(3) and~(4) is significant. If so,
and the LMC novae do in fact fade somewhat faster
for a given H$\alpha$ emission-line width compared with M31 novae,
the difference could be due to a lower metallicity (and hence opacity)
in the expanding shells of LMC novae that causes ejecta of all masses to become
optically thin sooner,
resulting in a population of novae with generally faster rates of decline.

\subsection{The Spatial Distribution of LMC Novae}

\begin{figure}
\includegraphics[angle=0,scale=.85]{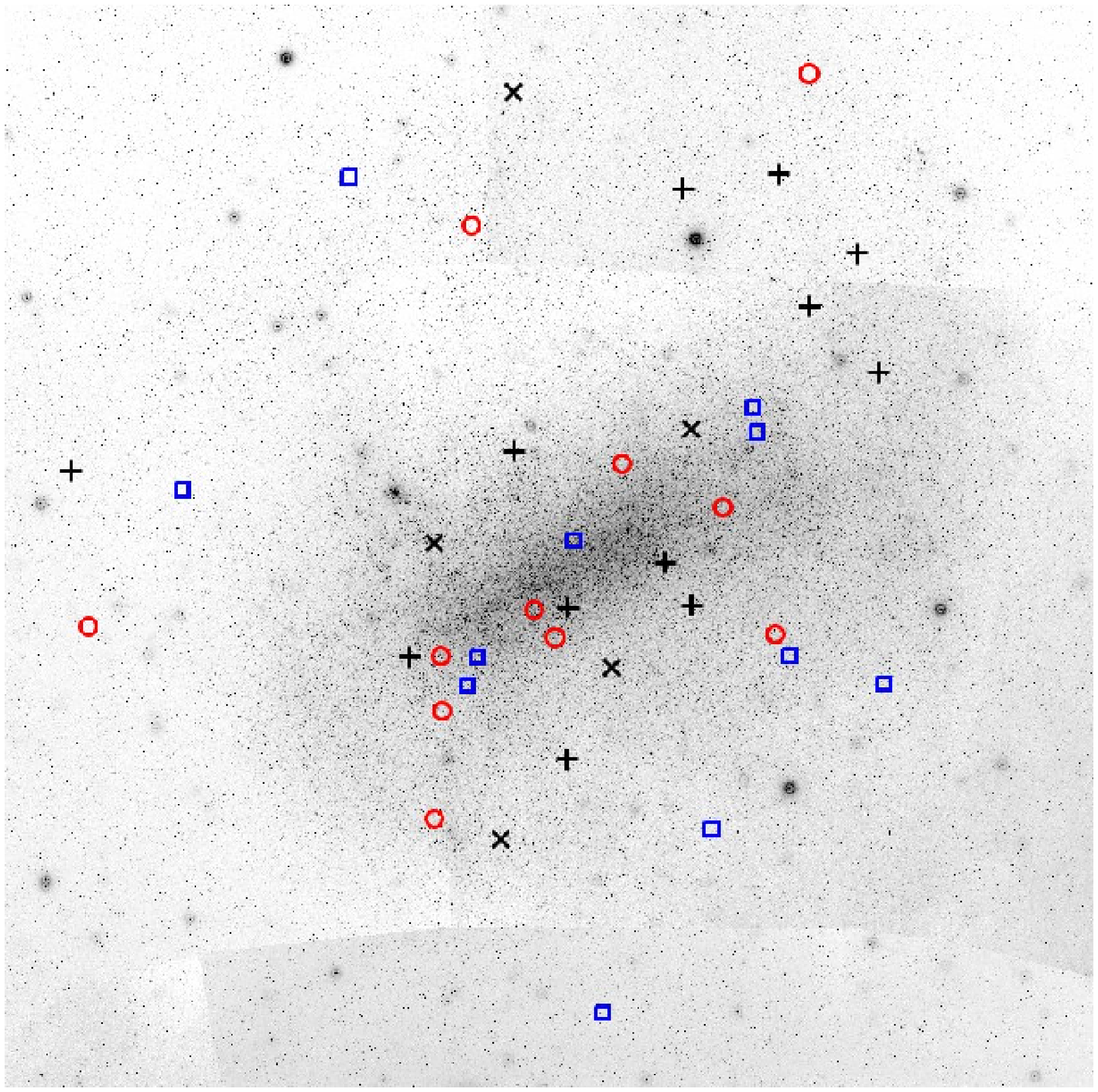}
\caption{The spatial distribution of the 40 observed nova candidates \citep
[LMC image from {\it Skyview},][]{mcg96}.
The Fe~II novae are
indicated by the red circles, while the
He/N and Fe~IIb (hybrid) novae are represented by the
blue squares. Novae with unknown spectroscopic types are represented
by ``+" symbols, with ``X" symbols representing questionable novae.
The novae do not appear to be strongly associated with the bar.
\label{fig1}}
\end{figure}

\begin{figure}
\includegraphics[angle=-90,scale=.65]{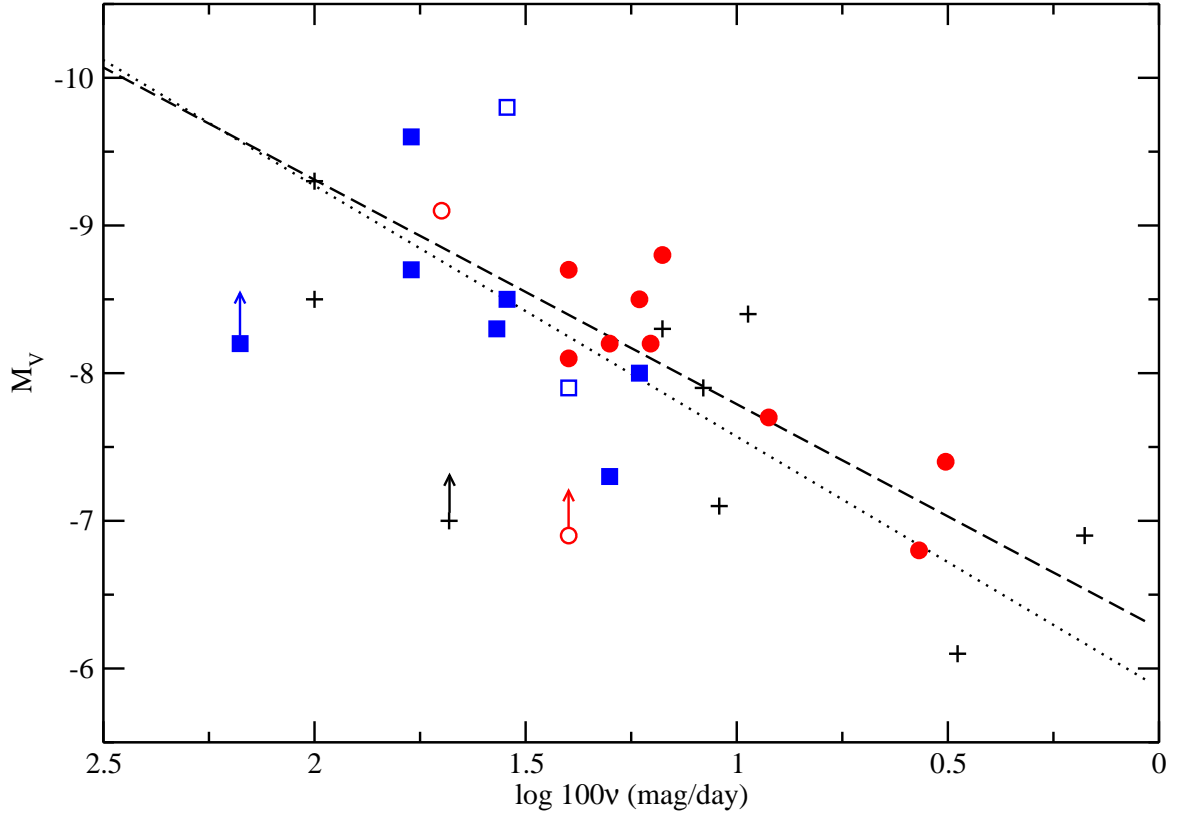}
\caption{The MMRD relation for LMC novae with measured fade rates.
Fe~II novae are represented by filled red circles, with
He/N and Fe~IIb novae shown as filled blue squares (open symbols
represent tentative spectroscopic types). Novae of
unknown spectroscopic type are shown as ``+" symbols. The three
novae with $M_V$ shown as upper limits are systems where maximum light
was missed, and no estimate of the peak brightness is available.
The dashed line represents a linear least-squares fit to the data
as given in Equation~(1). For comparison, the dotted line
(given by Equation~[2]) shows the MMRD
relation from \citet{sha11b} for their sample of M31 novae.
\label{fig2}}
\end{figure}

The LMC is an irregular galaxy of Hubble type Irr/SBr(s)m \citep{dev91}.
The luminosity profile has been the subject
of several studies, which have established the presence
of a barred disk seen near face-on,
with a brightness distribution that drops off exponentially from the center
of the galaxy. Figure~\ref{fig1} shows
the projected positions of the 40 known LMC
nova candidates from Table~\ref{novatable}
superimposed over an image of the galaxy. Of these 40,
spectroscopic classes are known or suspected for 21 of the novae.
To study their distributions,
we have plotted the Fe~II systems as red circles and
the He/N and Fe~IIb novae as blue squares.
The novae appear to be distributed uniformly across the face of the galaxy,
with no obvious dependence of spectroscopic class on spatial position. As
has been noted in previous studies \citep[e.g.,][]{sub02}, there
may be a slight enhancement of novae south east of the bar, with
a relative dearth in the 30 Dor region (north east of the bar). In addition,
as pointed out by \citet{van88}, despite the enhanced luminosity,
novae do not appear to be strongly associated with the central bar itself.

We can make a more quantitative study of the nova distribution
by comparing the nova density with that of the background light.
According to the photometry of \citet{dev72}, the disk of the
LMC is oriented at an inclination angle, $i=27^{\circ}$, with
a position angle, $PA=170^{\circ}$. Based on these parameters, we have
computed the projected distance of each nova from the center of the galaxy.
The radial light distribution is then computed assuming
an exponential disk with a central brightness of 21.16 mag~arcsec$^{-2}$, and
a scale length, $r_0=98.7'$ based on the $V$-band photometry of \citet{gal04}.
Figure~\ref{fig5} shows the resulting
cumulative nova distribution compared with the background $V$-band light.
The fit appears to be quite good, with
a Kolmogorov-Smirnov (KS) test indicating that the distributions
would be expected to differ by more than that
observed 52\% of the time if they were drawn
from the same parent population. Thus, there is no reason to
reject the hypothesis that the novae follow
the light distribution of the LMC.

\section{Discussion and Conclusions}

It has long been suggested that there exists two populations of novae,
a ``bulge" population and a ``disk" population, with the former
associated with an older bulge and thick disk population (Pop~II) and the
latter with a younger (Pop I), thin-disk component.
The distinction
is based upon limited observational data suggesting that bulge
novae are generally less luminous and have
slower photometric development when compared with their disk
counterparts \citep{due90,del92a}. There is also evidence to suggest that
the spectroscopic class of Galactic novae is affected by stellar population,
with He/N novae having a smaller distribution of
scale heights from the midplane of the Galactic disk \cite{del98}.
This finding is explained as the result of the
larger white dwarf masses in novae associated with
a younger disk population, which eject smaller accreted masses
at higher velocities compared with novae from older stellar populations.

There have been numerous studies of the relatively large
and equidistant sample of novae in the nearby spiral
galaxy M31, going all the way back to the early work of \cite{hub29}.
Recently, \citet{sha11b} considered a total of 91 M31 novae with
available spectroscopic data and found no evidence that
spectroscopic class was sensitive to spatial position within
the galaxy. They did, however, find that novae at larger galactocentric
radius faded slightly more quickly than novae closer to the center.
A caveat to their analysis is that the high inclination
of M31's disk to our line of sight makes it difficult to
clearly separate the disk and bulge component of the galaxy.
This is a particular problem in the central regions of the galaxy
(where most novae occur) where the foreground disk is projected
onto the central bulge.

Rather than attempting to disentangle
separate nova populations from a single galaxy like M31, it would
appear that a better approach to studying nova populations would be to
compare the properties of novae
in galaxies with differing Hubble types
and differing stellar populations. With that goal in mind \cite{sha12}
considered the spectroscopic classes and rates of decline for the
available sample of novae in M33, an
essentially bulge-less galaxy with a dominant disk population.
Although there were only 8 novae with known spectroscopic
class in M33 compared with the 91 available for M31 (where the
overall nova rate is $\sim10\times$ higher), there was already
sufficient data to suggest that the mix of spectroscopic classes was
dissimilar in the two galaxies at the 99\% confidence level.
In M31, approximately 80\% of the
novae are members of the Fe~II class, which is similar to what
is observed in the Galaxy \citep{sha07,wal12}. In M33, on the other hand,
Fe~II novae made up less than 40\% of the total. The limited light
curve data was insufficient to draw any conclusions regarding nova speed class.

\begin{figure}
\includegraphics[angle=-90,scale=.65]{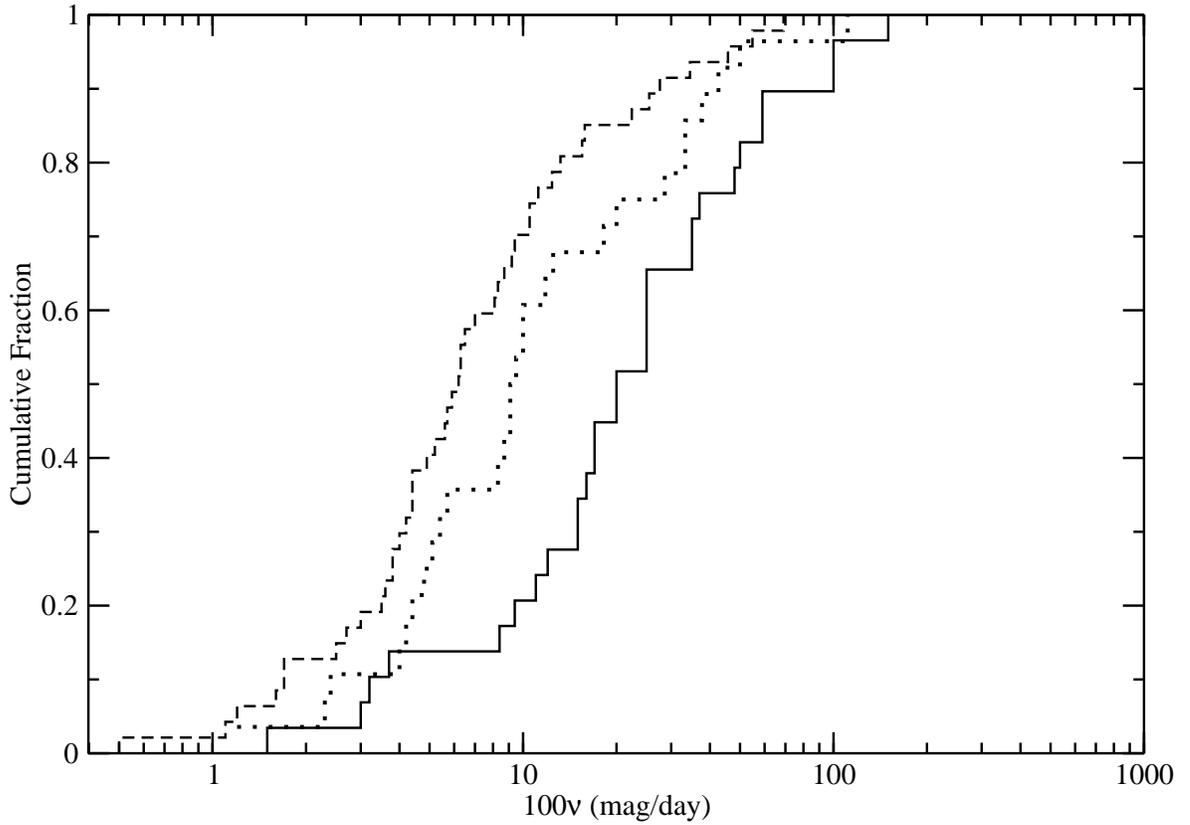}
\caption{The cumulative distribution of the fade rates for the
LMC novae compared with those of the Galaxy (dotted line)
and M31 (broken line).
The LMC novae are considerably ``faster" than their Galactic
and M31 counterparts.
A Kolmogorov-Smirnov test confirms that the distributions are markedly
different with KS$\sim$0.
\label{fig3}}
\end{figure}

\begin{figure}
\includegraphics[angle=-90,scale=.65]{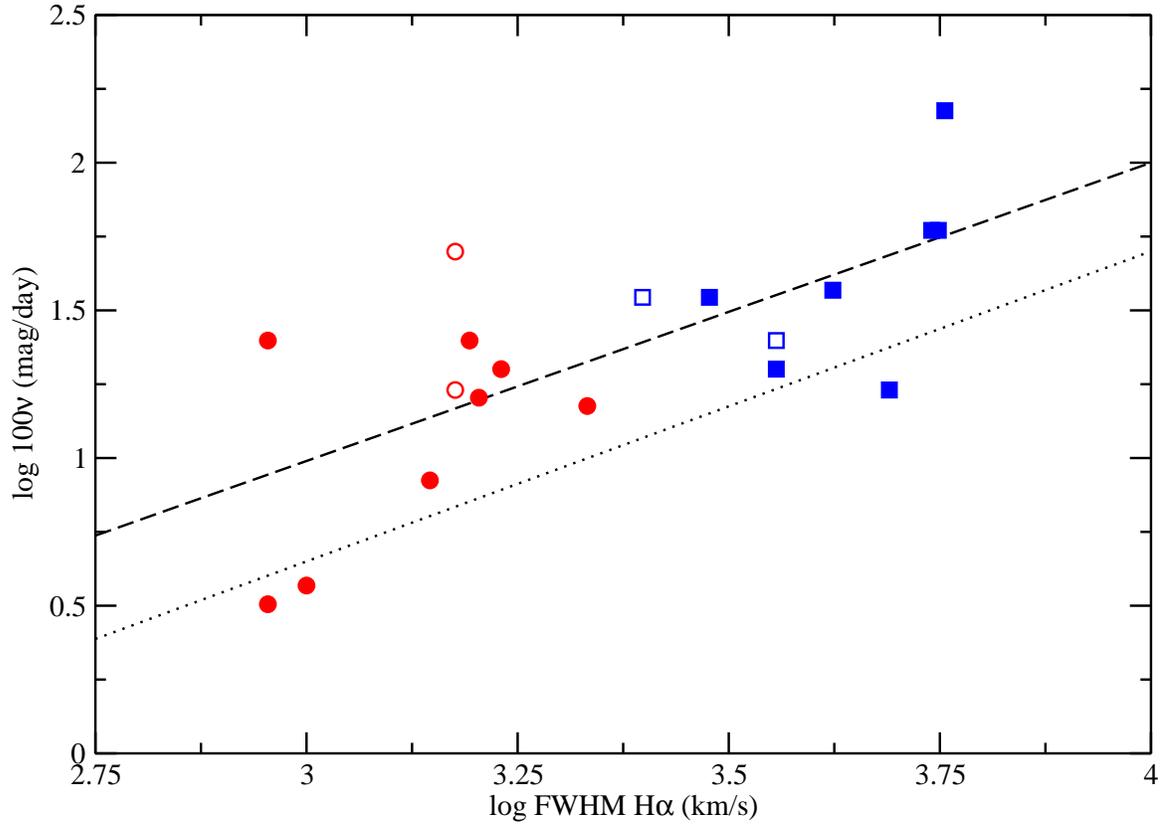}
\caption{The dependence of the
fade rate on nova expansion velocity (as reflected by
the FWHM of H$\alpha$). The symbols have the same meaning as in Figure~2.
There is a clear trend of increasing fade rate with
increasing H$\alpha$ emission line width.
The dashed and dotted lines reflect the best-fit relations given in the
text (Equations~[3] and~[4]) for the LMC and a sample of M31 novae
from \citet{sha11b}, respectively.
\label{fig4}}
\end{figure}

\begin{figure}
\includegraphics[angle=-90,scale=.65]{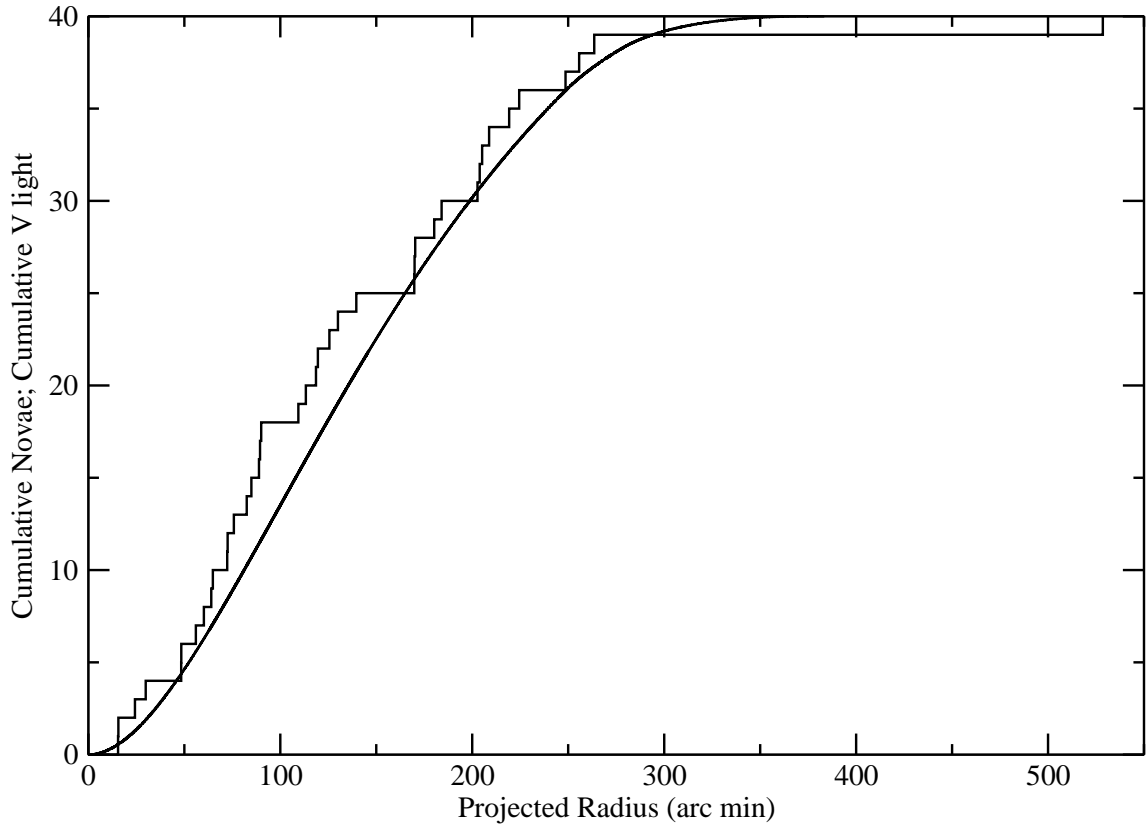}
\caption{The cumulative distribution of the LMC novae compared with the
cumulative distribution of the background $V$ light. The nova distribution
follows the light distribution well (KS = 0.52).
\label{fig5}}
\end{figure}

The only other galaxy for which a significant number of spectroscopic types
are available is the LMC. As we have seen, the fraction of
Fe~II novae in the LMC is approximately 50\%, which
appears to be significantly lower than that observed
in M31 and the Galaxy.
Following the analysis presented in \citet{sha12} for M33,
if we were to assume
the ratio of Fe~II to He/N (+ Fe~IIb) spectroscopic types is the
same as in M31, the probability of observing 9 or fewer
Fe~II novae in the LMC out of the 18 with known spectroscopic type would be
less than 0.5\%.
In other words, the mix of nova spectroscopic types in the LMC
differs from that observed in M31 at the 99.5\%
confidence level.
When compared with M31 (and the Galaxy),
He/N and Fe~IIb novae appear to
make up a significantly larger fraction of the nova population, not only in
M33, but in the LMC as well.
In addition to the differences in spectroscopic class, we have also
seen that the distribution of decline rates differs strikingly
between M31 and the LMC. As clearly seen in Figure~\ref{fig3},
LMC novae fade faster, on average, compared with novae in M31 and the Galaxy.

It seems unavoidable to conclude that the difference in dominant
stellar population between these galaxies is responsible for the
differences in their relative percentages of Fe~II novae and
in their average nova decline rates. Both of these
differences are likely the result of higher mean white dwarf masses
in LMC novae compared with novae in M31.
\citet{sub02} have studied the local (projected)
stellar populations in regions immediately surrounding 15 novae in the LMC.
Their principal conclusion is that most novae appear to arise from an
intermediate age population with star formation commencing
$\sim3\pm1$ Gyr ago. The fast and moderately fast novae appear to
arise from a somewhat younger population with mean age of
$1-3.2$ Gyr, while the one slow nova came from a region with
a stellar population having a wide range of ages going back to 10 Gyr.
Novae in M31 appear to come predominately from the
bulge~\citep{cia87,sha01,dar06}, which has an estimated age of
$>12$~Gyr~\citep{sag10}. As demonstrated by the population synthesis
models of \cite{yun97}, the younger stellar populations seen in
disk galaxies are expected to produce novae
at a higher rate compared with older stellar populations characteristic
of spiral bulges and elliptical galaxies. This difference
results from higher average white dwarf masses expected in recently formed
nova progenitor binaries~\citep[e.g.,][]{dek92,tut95,pol96}.
The higher white dwarf masses enable
a TNR to be triggered with a smaller accreted (and ejected) mass, resulting
not only in faster photometric evolutions and a higher He/N nova fraction,
but also in shorter average recurrence times and a higher nova rate.
Indeed, the LMC has the highest luminosity-specific nova rate measured
for any galaxy~\citep{del94,sha08}.

Finally, we note that, of the
35 confirmed nova systems in the LMC, 3 are recurrent. The fraction of RNe,
which approaches 10\% (or $\sim$16\% of the number of outbursts
recorded) appears to be significantly higher than that seen in
M31, where a total of only $\sim$8-15 RN systems (representing $\sim15-30$
eruptions) have been identified out of
a total of more than 900 nova eruptions cataloged.
Although these fractions are
certainly lower limits to the true fraction because of the
observational selection against the discovery of multiple outbursts, these
selection effects should affect the LMC and M31 observations
comparably. Thus, it appears likely
that the higher recurrent nova fraction observed in the LMC is indeed real.
Nova recurrence times are believed to be determined principally
by two parameters, the white dwarf's mass and its accretion rate \citep{tow05}.
Short recurrence times are predicted for novae having either
high white dwarf masses, high accretion rates, or both.
While it is true that among Galactic systems, RNe in 
the T~Pyx subclass appear to achieve their short
recurrence times as a result of high accretion rates rather than
from high white dwarf masses \citep[e.g.,][]{dar12}, it is
hard to understand why novae in the LMC would have systematically higher
accretion rates compared with novae in M31. Rather,
it would appear more plausible to attribute
the higher fraction of RNe in the LMC
to higher mean white dwarf masses associated with the younger population of
novae in this galaxy. This conclusion is further supported by
the fact that the RN candidates in the LMC are all members of the He/N
spectroscopic class, where systems harboring
massive white dwarfs are expected to dominate.

\acknowledgments

I thank B. Schaefer for discussions about LMC novae, 
and M. Bode, R. Ciardullo and M. Darnley
for comments on an early draft of the manuscript.
The use of NASA's SkyView facility
(http://skyview.gsfc.nasa.gov) located at NASA Goddard
Space Flight Center is gratefully acknowledged.
A.W.S. benefited from financial support through NSF grant AST-1009566.




\clearpage

\begin{deluxetable}{lcccrrlr}
\tabletypesize{\scriptsize}
\tablenum{1}
\tablewidth{0pt}
\tablecolumns{8}
\tablecaption{LMC Nova Candidates\label{novatable}}
\tablehead{\colhead{} & \colhead{JD Discovery}&\colhead{Discovery} & \colhead{} & \colhead{$\Delta\alpha~cos\delta$\tablenotemark{a}} & \colhead{$\Delta\delta$\tablenotemark{a}} & \colhead{} & \colhead{}  \\
\colhead{Nova} & \colhead{(2,400,000+)} & \colhead{mag} & \colhead{Filter} & \colhead{($'$)} & \colhead{($'$)} & \colhead{Type} & \colhead{References\tablenotemark{b}}}
\startdata
LMCN 1926-09a &  24786.5 & 12.4 &$pg$& $ -51.248$&$ 176.461$ & \dots & 1 \cr
LMCN 1935-09a &  28049.0 & 11.0 &$pg$& $-472.014$&$  37.481$ & \dots & 1 \cr
LMCN 1936-02a &  28223.5 & 10.8 &$pg$& $ -95.155$&$ 183.077$ & \dots & 1 \cr
LMCN 1937-11a &  28860.5 & 10.6 &$pg$& $ 182.642$&$  37.633$ & \dots & 1 \cr
LMCN 1948-12a &  32891.0 & 13.0 &$pg$& $  74.250$&$ -37.799$ & \dots & 1 \cr
LMCN 1951-01a &  33651.0 & 11.9 &$pg$& $ -55.260$&$ -14.069$ & \dots & 1 \cr
LMCN 1952\tablenotemark{c}     &    \dots & $<11.4$ &$V$ & $  26.354$&$ 220.441$ & \dots & 1 \cr
LMCN 1966\tablenotemark{c}     &    \dots & $<11.1$ &$V$ & $  32.518$&$-121.150$ & \dots & 1 \cr
LMCN 1968-12a &  40205.0 & 10.4 &$pg$& $ -63.910$&$-116.345$ & \dots & 1 \cr
LMCN 1970-03a &  40653.7 & 12.0 &$V$ & $  48.044$&$ -50.620$ & Fe IIb?& 1,2 \cr
LMCN 1970-11a &  40972.5 & 11.0 &$V$ & $  59.194$&$ -62.458$ & Fe II & 1,3 \cr
LMCN 1971-03a &  41039.5 & 11.8 &$V$ & $-141.268$&$  92.332$ & \dots & 1 \cr
LMCN 1971-08a &  41079.9 & 13.0 &$V$ & $ 101.185$&$ 181.798$ & \dots & 1 \cr
LMCN 1972-08a &  41551.5 & 11.0 &$pg$& $  26.467$&$  56.254$ & \dots & 1 \cr
LMCN 1973-09a &  41641.0 & 11.6 &$pg$& $ -43.113$&$   5.256$ & \dots & 1 \cr
LMCN 1977-02a &  43200.5 & 12.6 &$V$ & $ 229.730$&$  47.415$ & \dots & 1 \cr
LMCN 1977-03a &  43214.5 & 10.7 &$V$ & $ -93.622$&$ -27.194$ & Fe II & 1,4 \cr
LMCN 1978-03a &  43596.5 & 12.0 &$V$ & $-108.690$&$ 228.834$ & Fe II?& 1,5 \cr
LMCN 1978-11a &  43814.5 & 16.0 &$V$ & $-131.191$&$ 146.751$ & \dots & 1 \cr
LMCN 1981-09a &  44877.9 & 12.0 &$pg$& $  43.230$&$ -37.590$ & Fe IIb & 1,6 \cr
LMCN 1987-09a &  47056.0 &  9.6 &$V$ & $   1.352$&$ -15.379$ & \dots & 1 \cr
LMCN 1988-03a &  47242.0 & 11.4 &$V$ & $  60.054$&$ -37.593$ & Fe II & 1,7 \cr
LMCN 1988-10a &  47448.0 & 11.3 &$V$ & $ -85.001$&$  65.061$ & He/N  & 1,8 \cr
LMCN 1990-01a &  47908.0 & 11.5 &$V$ & $  -1.110$&$  15.560$ & He/N  & 1,9 \cr
LMCN 1990-02a\tablenotemark{d} &  47936.6 & 11.2 &$V$ & $ -64.203$&$-116.343$ & He/N  & 1,10,11,12,13 \cr
LMCN 1991-04a &  48362.5 &  9.0 &$V$ & $-100.150$&$ -36.940$ & Fe IIb? & 1,14,15 \cr
LMCN 1992-11a &  48940.1 & 10.7 &$R$ & $ -22.913$&$  50.589$ & Fe II & 1,16,17,18 \cr
LMCN 1995-02a &  49773.6 & 10.7 &$V$ & $  16.726$&$ -16.142$ & Fe II & 19 \cr
LMCN 1996\tablenotemark{c}     &    \dots & $<12.4$ &$V$ & $ -55.060$&$  66.248$ & \dots & 1 \cr
LMCN 1997-06a &  50615.9 & $<12.7$ &$V$ & $-109.142$&$ 122.573$ & \dots & 1 \cr
LMCN 1998-12a\tablenotemark{c} &  51176.2 & 16.9 &$R$ & $  62.831$&$  13.962$ & \dots & 1 \cr
LMCN 1999\tablenotemark{c}     &    \dots & 12.5 &$V$ & $ -18.432$&$ -42.775$ & \dots & 1 \cr
LMCN 2000-07a &  51737.9 & 11.2 &$V$ & $   7.320$&$ -28.939$ & Fe II & 1,20 \cr
LMCN 2001-06a &  52067.5 & $<9.7$ &$R$ & $   2.059$&$ -84.652$ & \dots & 1 \cr
LMCN 2002-02a &  52332.6 & 10.5 &$V$ & $  62.542$&$-111.955$ & Fe II & 1,18 \cr
LMCN 2003-06a &  52808.5 & $<11.0$ &$V$ & $ -83.487$&$  76.441$ & He/N? & 21 \cr
LMCN 2004-10a\tablenotemark{e} &  53298.7 & 10.8 &$V$ & $ 178.462$&$  38.732$ & He/N  & 1,22,23 \cr
LMCN 2005-09a &  53644.0 & 12.0 &$V$& $ 221.767$&$ -23.805$ & Fe II??& 1,24 \cr
LMCN 2005-11a &  53696.6 & 12.8 &$W$ & $ -69.355$&$  30.929$ & Fe II & 1,25,26 \cr
LMCN 2009-02a\tablenotemark{f} &  54867.6 & 10.6 &$W$ & $ 102.039$&$ 181.838$ & He/N  & 1,26,27 \cr
LMCN 2009-05a &  54956.5 & 12.1 &$W$ & $  45.964$&$ 159.098$ & Fe II & 1,26 \cr
\tablebreak
LMCN 2012-03a &  56012.9 & 10.7 &$W$ & $-143.471$&$ -50.035$ & He/N  & 1,26,28 \cr
LMCN 2012-10a &  56225.8 & 11.5: &$I$ &  $-14.084$ & $-200.677$ & He/N  & 1,26,29,30 \cr
\enddata
\tablenotetext{a}{Offsets from the center of the LMC ($\alpha_\mathrm{J2000} = 5^h23^m34\fs50, \delta_\mathrm{J2000} = -69\degr45\arcmin22\farcs0$)}
\tablenotetext{b}{References:
(1) Pietsch {\tt (http://www.mpe.mpg.de/\~{}m31novae/opt/lmc/)};
(2) \citet{mac70};
(3) \citet{hav72};
(4) \citet{can81};
(5) \citet{gra79};
(6) \citet{maz81};
(7) \citet{sch98};
(8) \citet{sek89};
(9) \citet{dop90};
(10) \citet{wil90};
(11) \citet{sek90a};
(12) \citet{sho91};
(13) \citet{sek90b};
(14) \citet{del91b};
(15) \citet{sch01};
(16) \citet{del92};
(17) \citet{due92};
(18) \citet{mas05};
(19) \citet{del95b};
(20) \citet{due00};
(21) H. Bond (2012, private communication);
(22) \citet{bon04};
(23) \citet{mas04};
(24) \citet{rea09};
(25) \citet{wal05};
(26) \citet{wal12};
(27) \citet{ori09};
(28) \citet{pri12};
(29) \citet{wyr12}.
(30) \citet{pri13};
}
\tablenotetext{c}{Poorly known object}
\tablenotetext{d}{Recurrent nova candidate: LMCN 1968-12a}
\tablenotetext{e}{Recurrent nova candidate: LMCN 1937-11a (YY Dor)}
\tablenotetext{f}{Recurrent nova candidate: LMCN 1971-08a}
\end{deluxetable}

\clearpage

\begin{planotable}{lcccccccllr}
\tabletypesize{\scriptsize}
\tablenum{2}
\tablewidth{0pt}
\tablecolumns{11}
\tablecaption{Light Curve Parameters\label{lcparam}}
\tablehead{\colhead{Nova} & \colhead{Filter} & \colhead{$m_{\mathrm{max}}$} &\colhead{$\Delta t$ (d)\tablenotemark{a}}& \colhead {$E(B-V)$} & \colhead{$M_{\mathrm{max}}$\tablenotemark{b}} & \colhead{$t_2$ (d)} & \colhead{$t_3$ (d)} & \colhead{$\nu$ (d$^{-1}$)} & \colhead{Type} & \colhead{References\tablenotemark{c}}}
\startdata
LMCN 1926-09a & $pg$ & 12.0 &\dots &\dots & $ -6.9$  &    \dots       &  $  200 $ & 0.015 & \dots & 1,2 \\
LMCN 1935-09a & $pg$ & 11.0 &\dots &\dots & $ -7.9$  &    \dots       &  $  25  $ & 0.12 & \dots & 1,2 \\
LMCN 1936-02a & $pg$ & 10.5 &\dots &\dots & $ -8.4$  &    \dots       &  $  32  $ & 0.094 & \dots & 1,2 \\
LMCN 1937-11a & $pg$ & 10.6 &\dots &\dots & $ -8.3$  &    \dots       &  $  20  $ & 0.15 & \dots & 1,2 \\
LMCN 1948-12a & $pg$ & 13.0 &\dots & 0.18 & $ -6.1$: &    \dots       &  $  100 $ & 0.030 & \dots & 1,2 \\
LMCN 1951-01a & $pg$ & 11.9: &\dots &\dots & $ <-7.0$:\tablenotemark{d}  &    \dots       &  $  6.3 $ & 0.48 & \dots & 1,2 \\
LMCN 1968-12a & $pg$ & 10.4: &1&\dots & $ -8.5:$  &    \dots       &  $  5.26 $& 1.0: & \dots & 3,4,5 \\
LMCN 1970-11a & $V$ &  10.8: &7&\dots & $ -8.1$:  &    \dots    &  \dots & 0.25: & Fe II & 4,5,6 \\
LMCN 1971-03a & $V$ &  11.8 &7&\dots & $ -7.1 $  &    \dots       &  $  28.3 $& 0.11 & \dots & 4,5,6 \\
LMCN 1977-03a & $V$ &  10.7 &0&0.11 & $ -8.2 $  &  $  11       $  &  21 & 0.16 & Fe II &5,7,8 \\
LMCN 1978-03a & $V$ &  9.8: &10 &\dots & $ -9.1$:  &  \dots     &  $  7.8  $& 0.50 & Fe II?& 4,5,9 \\
LMCN 1987-09a & $V$ &  9.6 &0 &\dots &  $ -9.3 $  &    \dots       &  $  5.3  $& 1.0 & \dots & 4,5,10 \\
LMCN 1988-03a & $V$ &  11.2&2&0.15 &  $ -7.7 $  &  $  22.5\pm4  $  &  $ 38.4^{+5.9}_{-5.0} $& 0.084 & Fe II & 11 \\
LMCN 1988-10a & $V$ &  10.3 &0&0.079 & $ -8.5 $  &  $  5         $  &  $  10   $& 0.35 & He/N & 12 \\
LMCN 1990-01a & $V$ & 9.7: &3&0.22 & $ -9.6$:  &4.5  &\dots& 0.59 & He/N & 13,14,15 \\
LMCN 1990-02a\tablenotemark{e} & $V$ & 10.2: &11 &\dots & $ -8.7$:    &  3  &\dots& 0.59: & He/N & 4,13,15,16 \\
LMCN 1991-04a & $V$ & 9.0 &0&0.10 & $ -9.8 $  &  $  6\pm1     $  &  $  8\pm1$& 0.35 & Fe IIb? & 15,17 \\
LMCN 1992-11a & $V$ & 10.2 &0&\dots & $ -8.7 $  &  $  6.9\pm1.1 $  &  $13.7\pm1.6$& 0.25 & Fe II & 11,15 \\
LMCN 1995-02a & $V$ & 10.4 &4&\dots & $ -8.5 $  &  $ 11.0\pm3.0 $  &  $19.6\pm3.2$& 0.17 & Fe II & 11,15 \\
LMCN 2000-07a & $V$ & 10.7:  &13 &\dots & $ -8.2$:  &  $ 8.0^{+4.5}_{-3.5} $  &  $20.0^{+12.0}_{-7.0}$& 0.20 & Fe II & 11,18 \\
LMCN 2002-02a & $V$ & 10.1 &4&\dots & $ -8.8 $  &  $ 12.0       $  &  $   23.0  $& 0.15 & Fe II & 4,13,15 \\
LMCN 2003-06a & $V$ & 11.0: &16 &\dots & $ -7.9$: & \dots     &  \dots& 0.25  & He/N? & 15,19 \\
LMCN 2004-10a\tablenotemark{f} & $V$ & 10.9: &3 &\dots & $ -8.0$:  &  \dots   &  \dots & 0.17 & He/N & 15 \\
LMCN 2005-09a & $V$ & $<12.0$ &\dots&\dots & $ <-6.9 $  &  8:  &  \dots & 0.25: & Fe II?? & 20 \\
LMCN 2005-11a & $V$ & 11.5: &0 &\dots & $ -7.4:  $  &  63 &  94  & 0.032 & Fe II &  21 \\
LMCN 2009-02a\tablenotemark{g} & $W$ & 10.6: &5 &\dots & $ -8.3$:  &  \dots  &  8: & 0.37: & He/N & 22,23 \\
LMCN 2009-05a & $W$ & 12.1: &11 &\dots & $ -6.8$:  & \dots &\dots& 0.037: & Fe II & 22,24 \\
LMCN 2012-03a & $W$ & $<10.7$: &12 &\dots & $<-8.2$:  &  1.1:& 2.1: & 1.5: &  He/N &22,25 \\
LMCN 2012-10a & $I$ & 11.5: &3&\dots & $ -7.3$:  &  10 & 15 & 0.20 &  He/N &26 \\
\enddata
\tablenotetext{a}{Upper limit on time elapsed between outburst maximum and discovery.}
\tablenotetext{b}{$\mu_o=18.50$ \citep{fre01} and
$E(B-V)=0.12$ \citep{ima07} assumed, unless otherwise noted.}
\tablenotetext{c}{References:
(1) \citet{hen54};
(2) \citet{bus55};
(3) \citet{sie70};
(4) \citet{sub02};
(5) \citet{cap90};
(6) \citet{gra71};
(7) \citet{can77};
(8) \citet{can81};
(9) \citet{gra78b};
(10) \citet{mcn87};
(11) \citet{hea04};
(12) \citet{sek89};
(13) \citet{mas05};
(14) \citet{van99};
(15) \citet{lil05a};
(16) \citet{sek90b};
(17) \citet{sch01};
(18) \citet{gre03};
(19) \citet{lil03};
(20) \citet{rea09};
(21) \citet{lil07};
(22) \citet{wal12};
(23) \citet{lil09a};
(24) \citet{lil09b};
(25) \citet{lil12};
(26) \citet{wyr12}.}
\tablenotetext{d}{Heavily extincted \citep{bus55}}
\tablenotetext{e}{Recurrent nova: LMCN 1968-12a}
\tablenotetext{f}{Recurrent nova candidate: LMCN 1937-11a (YY Dor)}
\tablenotetext{g}{Recurrent nova candidate: LMCN 1971-08a}
\end{planotable}

\begin{planotable}{lllclr}
\tabletypesize{\scriptsize}
\tablenum{3}
\tablewidth{0pt}
\tablecolumns{6}
\tablecaption{H$\alpha$ Emission-Line Properties\label{balmerline}}
\tablehead{\colhead{} & \colhead{FWHM} & \colhead{$\nu$} & \colhead{Phase\tablenotemark{a}} & \colhead{} & \colhead{} \\ \colhead{Nova} & \colhead{(km~s$^{-1}$)}  & \colhead{(d$^{-1}$)} & \colhead{(d)} & \colhead{Type} & \colhead{References\tablenotemark{b}}}
\startdata
LMCN 1970-11a &   $ 1560\tablenotemark{c}   $ &0.25: & 9,10,11 & Fe II & 1 \cr
LMCN 1977-03a &   $ 1600   $ &0.16 & 1 & Fe II & 2 \cr
LMCN 1978-03a &   $ 1500\tablenotemark{c}  $ &0.50& 10:& Fe II? & 3 \cr
LMCN 1981-09a &   $ 3800   $ & \dots & 1 & Fe IIb & 4 \cr
LMCN 1988-03a &   $ 1400\tablenotemark{d}   $ & 0.084 & 4,17,31 & Fe II & 5,6 \cr
LMCN 1988-10a &   $ 3000\tablenotemark{e}   $ & 0.35& 1,19,49 &  He/N  & 7 \cr
LMCN 1990-01a &   $ 5600   $ & 0.59 & 7,16 & He/N & 8 \cr
LMCN 1990-02a\tablenotemark{f} &   $ 5500   $ & 0.59:& 8 & He/N  & 9,10 \cr
LMCN 1991-04a &   $ 2500   $ & 0.35 &18 & Fe IIb? & 11 \cr
LMCN 1992-11a &   $ 900\tablenotemark{c}    $ & 0.25 & 1 & Fe II & 12,13 \cr
LMCN 1995-02a &   $ 1500\tablenotemark{c}   $ & 0.17 & 1 & Fe II? & 14 \cr
LMCN 2000-07a &   $ 1700   $ & 0.20 & 3 & Fe II & 15 \cr
LMCN 2002-02a &   $ 2150\tablenotemark{e}   $ & 0.15 & 6,11 & Fe II & 16 \cr
LMCN 2003-06a &   $ 3600   $ & 0.25 & 11& He/N? & 17 \cr
LMCN 2004-10a\tablenotemark{g} &   $ 4900\tablenotemark{e}   $ & 0.17 & 1,5 & He/N  & 18,19 \cr
LMCN 2005-11a &    900   & 0.032 & 2 & Fe II & 20,22 \cr
LMCN 2009-02a\tablenotemark{h} &  $ 4200   $ & 0.37: & 3 & He/N  & 21,22 \cr
LMCN 2009-05a &    1000   & 0.037:  & 7 & Fe II & 22 \cr
LMCN 2012-03a &   $ 5700   $ & 1.5: & 1 & He/N  & 23 \cr
LMCN 2012-10a &   $ 3600   $ & 0.20 & 10,16& He/N  & 22 \cr
\enddata
\tablenotetext{a}{Time elapsed between maximum light and spectra used for classification.}
\tablenotetext{b}{References:
(1) \citet{hav72};
(2) \citet{can81};
(3) \citet{gra79};
(4) \citet{maz81};
(5) \citet{dre90};
(6) \citet{sch98};
(7) \citet{sek89};
(8) \citet{dop90};
(9) \citet{sek90b};
(10) \citet{sho91};
(11) \citet{del91b};
(12) \citet{del92};
(13) \citet{due92};
(14) \citet{del95b};
(15) \citet{due00};
(16) \citet{mas05};
(17) H. Bond (2012, private communication);
(18) \citet{bon04};
(19) \citet{mas04};
(20) \citet{wal05};
(21) \citet{ori09};
(22) \citet{wal12};
(23) \citet{pri12}.}
\tablenotetext{c}{Velocity based on displacement of P Cyg absorption.}
\tablenotetext{d}{Mean of values from \citet{dre90} and \citet{sch98}}
\tablenotetext{e}{FWHM estimate assumed to be half of the FWZI.}
\tablenotetext{f}{Recurrent nova: LMCN 1968-12a}
\tablenotetext{g}{Recurrent nova candidate: LMCN 1937-11a (YY Dor)}
\tablenotetext{h}{Recurrent nova candidate: LMCN 1971-08a}
\end{planotable}

\end{document}